\documentclass[12pt]{article}
\usepackage[dvipdfmx]{graphicx}
\usepackage{float}
\usepackage{amsmath,amssymb}
\begin{document}
\global\arraycolsep=2pt
\newcommand{\p}{{\bf p}}
\newcommand{\q}{{\bf q}}
\newcommand{\s}{{\bf s}}
\newcommand{\x}{{\bf x}}
\newcommand{\y}{{\bf y}}
%
%
%
\thispagestyle{empty} 
\begin{titlepage}    
\begin{flushright}
TNCT-2101
\end{flushright}
\vspace{1cm}
\topskip 3cm
\begin{center}
  \Large{\bf
    Vacuum angle theta in a strongly interacting hidden sector
     and its effect
                }  \\
\end{center}                                   
\vspace{0.6cm}
\begin{center}
Shinji Maedan            
  \footnote{ E-mail: maedan@tokyo-ct.ac.jp}   
           \\
\vspace{0.8cm}
  \textsl{ Department of Physics, Tokyo National College of Technology,
               Kunugida-machi,Hachioji-shi, Tokyo 193-0997, Japan}
  \end{center}                                               
\vspace{0.5cm}
\begin{abstract}
\noindent       
\\
Introducing a vacuum angle $\theta_h$ in a strongly interacting hidden sector like QCD,
   we study how $\theta_h$ affects physical quantities in a scale
   invariant extension of the standard model with that hidden sector.
In this model the dynamical chiral symmetry breaking occurs in the hidden
   sector and generates a scale $ \vert \langle {\bar q} q \rangle \vert $ which triggers
   electroweak symmetry breaking.
We find that the expression for a vacuum expectation value of the Higgs field depends on
   $\theta_h$, while the condition which determines the Higgs particle mass does not.
This model contains the hidden-sector pions $\pi^0$ and $\pi^{\pm}$ which are candidates
   for cold dark matter, and we find the mass difference between the hidden-sector
   $\pi^0$ and $\pi^{\pm}$ will not be negligible for $\theta_h \approx \pi$.
\end{abstract} 
\end{titlepage}
%
%
%
%
\section{Introduction}
Attempts of scale invariant extension of the standard model (SM) are significant
   because there arises the hierarchy problem in many models beyond the SM.
In Ref.\cite{rf:HurKo}, Hur and Ko proposed a scale invariant extension of the SM with
   a new QCD-like strong interaction with $ N_{h,c}=3 $ colors in the hidden sector.
In their extension, the dynamical chiral symmetry breaking occurs in the hidden sector
   and its dynamical scale $ {\vert \langle {\bar q} q \rangle \vert}^{1/3} \sim O( {\rm TeV} )$
   is transmitted to the Higgs sector in SM by a real singlet scalar $S$ introduced.
This $S$ connects the hidden sector and the SM, triggering the electroweak (EW) symmetry
   breaking.
Their model also includes candidates of cold dark matter (CDM), i.e., the lightest mesons
   and baryons in the hidden sector.
In Ref.\cite{rf:HurKo}, however, a vacuum angle $\theta_h$ was not considered in the
   hidden sector with a new QCD-like strong interaction.

In this paper we shall consider the vacuum angle $\theta_h$ in the hidden
   sector\cite{rf:AntRedStrVig} of this model\cite{rf:HurKo} and study
   its effects on physical quantities such as the Higgs field and hidden-sector mesons.
Before dealing with $\theta_h$ in the QCD-like strong interaction (hidden sector), it would
   be helpful to recall the vacuum angle $\theta$ in ordinary QCD \cite{rf:VicPan}.
The rich structure of QCD vacuum gives rise to the following effective Lagrangian term,
\begin{equation}
   \frac{\theta}{16 \pi^2} \, {\rm tr} F_{\mu \nu} {\tilde F^{\mu \nu} },
  \label{aa}
\end{equation}
where $\theta$ is called a vacuum angle in QCD and this term violates P and CP
   conservation.
In order to investigate $\theta$ dependence in QCD, effective models such as the
   Di-Vecchia-Veneziano model have been widely
   used\cite{rf:RosSchTra,rf:VecVen,rf:KawOht,rf:NatArn}.
Di Vecchia and Veneziano constructed an effective Lagrangian\cite{rf:VecVen}
   in the large- $N_{color}$ limit of QCD to resolve the U(1) problem\cite{rf:Wit}.
The Di-Vecchia-Veneziano model is a low-energy effective theory of Nambu-Goldstone (NG)
   bosons $ \pi_{i} (i=1,2, \cdots, 8) $
   including the flavor-singlet pseudoscalar particle $\eta^0$, which will mix with
    $ \pi_8 $ and  $ \pi_3 $ to form the mass eigenstates $\eta', \eta$, and  $ \pi^0 $.
While authors in Ref.\cite{rf:VecVen} studied their effective theory for arbitrary
   $\theta$, the experimental upper limit on an electric dipole moment for the neutron or the
   CP-violating decay of NG boson $ \eta \rightarrow \pi^+  \, \pi^- $ bounds the $\theta$
   in QCD, $ \vert \theta \vert < 10^{-10} $.
Why the vacuum angle $\theta$ in QCD is nearly zero, and this is the strong CP problem
   which is not solved yet.

Returning to the discussion of the scale invariant extension of the SM\cite{rf:HurKo},
   we shall consider a vacuum angle $\theta_h$ in the hidden sector with a new
   QCD-like strong interaction.
Contrary to the case of QCD, there is no reason to restrict the value of $\theta_h$
   in the hidden sector because no particle in the hidden sector is detected in experiment.
As a low-energy effective theory of the hidden sector, we use the
   Di-Vecchia-Veneziano model with two flavors, containing the hidden-sector pions
   $ {\vec \pi} =(  \pi_1, \pi_2, \pi_3 ) $, the hidden-sector flavor-singlet
   pseudoscalar $ \eta^0 $, and the hidden-sector vacuum angle $\theta_h$.
In the model of Ref.\cite{rf:HurKo}, as noted before, the SM sector is connected with
   the hidden sector by the messenger $S$ (a real scalar).
Since the ground state energy of the Di-Vecchia-Veneziano Lagrangian in the hidden sector
   depends on $\theta_h$, the Higgs field will be affected by $\theta_h$,
   which we shall study.
Other effects of $\theta_h$ on physical quantities will be seen, for example,
   in the CP-violating interactions involving the hidden-sector
   pions\cite{rf:RedStrTesVig}, or in the mass difference of the hidden-sector isospin multiplet,
   i.e., the hidden-sector pions that are candidates of CDM.
We examine how the mixing term between  $ \pi_3 $ and $ \eta^0 $ depends on
   $\theta_h$, and calculate the mass difference between the hidden-sector
   $ \pi^0 $ and $ \pi^\pm $.

This paper is organized as follows.
In Sect.2, as a low-energy effective theory of the hidden sector the Di-Vecchia-Veneziano
   model with two flavors is introduced and its ground state energy depending on
   a vacuum angle $\theta_h$ is presented\cite{rf:VecVen,rf:Wit2}.
In Sect.3, we discuss physical effects of $\theta_h$ on the Higgs field in SM sector and
   also on the mass difference between the hidden-sector pions $ \pi^0 $
   and $ \pi^\pm $, which are dark mater candidates.
The last section is devoted to the conclusion.
%
%
%
%
%
%
\section{Ground state in a strongly interacting hidden sector
                with vacuum angle}
The hidden sector Lagrangian\cite{rf:HurKo} with the vacuum angle $\theta_h$ is given by
\begin{equation}
    {\cal L_{\rm H} } = - \frac{1}{2} {\rm tr} \, F_{\mu \nu} F^{\mu \nu}
          + \sum_{k=1}^{N_{h,f}} {\rm tr} \, {\bar q}_k ( i \gamma^\mu D_\mu - y_k S ) q_k
          + \frac{\theta_h}{16 \pi^2} \, {\rm tr} F_{\mu \nu} {\tilde F^{\mu \nu} },        
  \label{ba}
\end{equation}
where $F_{\mu \nu}$ is the field strength for a hidden $SU(N_{h,c})$ gauge sector and
   $q$ the hidden-sector quark transforming as a fundamental representation of
   $SU(N_{h,c})$.
We choose the number of flavors $N_{h,f}$ to be two, $N_{h,f}=2$.
A real singlet scalar $S$ is introduced in this Lagrangian as a messenger
   between the hidden sector and SM sector.
The scalar potential part for
   the Higgs doublet field $H$ of the SM Lagrangian is modified
   as follows\cite{rf:HurKo,rf:HolKubLimLin}:
\begin{equation}
   V_{ {\rm SM}+ S } = \lambda_H ( H^\dagger H )^2 + \frac{1}{4} \lambda_S S^4
       -  \frac{1}{2}  \lambda_{HS} \, S^2 ( H^\dagger H ).
  \label{bb}
\end{equation}
Here we assume $ \lambda_{HS}>0 $.
For the stability of this potential with $ H =(0, h/ \sqrt{2})^T $, it is required that
    $ \lambda_H >0,  \lambda_S >0,$ and $ 4 \lambda_H  \lambda_S - \lambda_{HS}^2 >0$.
With this modified SM Lagrangian, ${\cal L}_{ {\rm SM}+ S }$, the total Lagrangian is
   $ {\cal L}_T = {\cal L}_H + {\cal L}_{ {\rm SM}+ S } $.
In the hidden sector with strong interaction the hidden-sector quarks will
   condensate, $ \langle {\bar q} q \rangle \ne 0 $,
   by non-perturbative effect and this scale $ \langle {\bar q} q \rangle $ will be transmitted
   to the Higgs sector in SM by the real scalar $S$.
If the field $S$ gets a vacuum expectation value (VEV), $ \langle S \rangle \ne 0 $,
   the hidden-sector quarks will obtain
   its mass, $ m_1 = y_1 \langle S \rangle, m_2 = y_2 \langle S \rangle $,
   through the Yukawa coupling in $\cal L_{\rm H}$\cite{rf:HurKo}.
We will make use of the Di-Vecchia-Veneziano model\cite{rf:VecVen} as a low-energy
   effective theory of the hidden-sector Lagrangian ${\cal L}_{\rm H}$
   with the hidden-sector quark mass, $m_1$ and $m_2$,
\begin{eqnarray}
    {\cal L}_{\rm DVV}
  &=& \frac{f_\pi^2}{4} \, {\rm Tr} \{ \partial_\mu U \partial^\mu U^\dagger \}
          + \frac{ \vert \langle {\bar q} q \rangle \vert }{2}  \,
                {\rm Tr} \{ {\cal M} U^\dagger + U {\cal M}  \}                         \nonumber \\
  & &  \hskip1cm - \frac{\tau}{2} \biggl\{ \theta_h + \frac{i}{2} \left(
            \log \det U - \log \det U^\dagger \right) \biggr\}^2,
  \label{bc}
\end{eqnarray}
where
\begin{equation}
   U = \exp \biggl[  \frac{i}{f_\pi} \biggl\{ \sum_{j=1}^{3} \pi_j \sigma^j
           + \eta^0 \cdot \bf 1 \biggr\}  \biggr],
  \label{bd}
\end{equation}
and
\begin{equation}
 {\cal M}  =  \left(   \begin{array}{cc}
                     m_1    &        0            \\
                     0         &       m_2   
                                         \end{array}
                                                        \right).
  \label{be}
\end{equation}
The $ \pi_{j} (j=1,2,3) $ are the hidden-sector pions with decay constant $f_\pi$,
   $\eta^0$ the hidden-sector flavor-singlet
   pseudoscalar meson, $ \vert \langle {\bar q} q \rangle \vert $ the absolute value of
   the hidden-sector quark condensate in the massless theory, and $\tau$ the topological
   susceptibility of the pure $SU(N_{h,c})$ Yang-Mills theory\cite{rf:VicPan}.

For the ground state of this system, $U$ takes the form
\begin{equation}
   U_g =  \left(   \begin{array}{cc}
                     e^{i \varphi_1}    &        0                     \\
                     0                   &         e^{i \varphi_2}
                                         \end{array}
                                                        \right),
  \label{bf}
\end{equation}
where $ \varphi_1$ and $ \varphi_2 $ can be determined modulo $2 \pi$.
The ground state energy $E_{\rm DVV} (\varphi_1, \varphi_2 )$ for
   $ {\cal L}_{\rm DVV} $ becomes
\begin{equation}
  E_{\rm DVV} (\varphi_1, \varphi_2 ) 
       = -  \vert \langle {\bar q} q \rangle \vert \left( m_1 \cos \varphi_1
            + m_2 \cos \varphi_2 \right) + \frac{\tau}{2} \left\{ \theta_h
                - ( \varphi_1 +\varphi_2 ) \right\}^2,
  \label{bg}
\end{equation}
and the angles $ \varphi_1$ and $ \varphi_2 $ should satisfy
\begin{eqnarray}
     m_1  \sin \varphi_1 &=&
     \biggl( \frac{\tau}{  \vert \langle {\bar q} q \rangle \vert } \biggr)
         \biggl\{ \theta_h -  ( \varphi_1 +\varphi_2 ) \biggr\},                     \nonumber \\
     m_2  \sin \varphi_2 &=&  m_1  \sin \varphi_1.
     \label{bh}
\end{eqnarray}
We shall consider the region of $\theta_h$, $ -\pi < \theta_h < \pi $.
Now let us assume that the hidden-sector quark mass
   $m_j (j=1,2)$ is small,
\begin{equation}
    m_j \ll 
    \biggl( \frac{\tau}{ \vert \langle {\bar q} q \rangle \vert } \biggr),
            \hskip1cm  (j=1,2),
  \label{bi}
\end{equation}
throughout this paper.
Under this assumption the ground state energy $ E_{\rm DVV} $ can be represented
   explicitly by $\theta_h$\cite{rf:VecVen,rf:Wit2,rf:MetZhi} as discussed below.
We seek solutions having the following form,
\begin{equation}
       \varphi_1 = \frac{\theta_h}{2} + \alpha
            + O  \biggl( \frac{ \vert \langle {\bar q} q \rangle \vert m_j}{\tau}\biggr),    \hskip1cm
        \varphi_2 =  \frac{\theta_h}{2} - \alpha
            + O  \biggl( \frac{ \vert \langle {\bar q} q \rangle \vert m_j}{\tau}\biggr).
  \label{bj}
\end{equation}
In the limit, 
   $ {\vert \langle {\bar q} q \rangle \vert m_j} / \tau \rightarrow 0 $,
   the $ \alpha $ should satisfy
\begin{equation}
   m_1 \sin  \biggl( \frac{\theta_h}{2} + \alpha \biggr)
   = m_2 \sin  \biggl( \frac{\theta_h}{2} - \alpha \biggr),
  \label{bk}
\end{equation}
and a solution of this equation was obtained by
\begin{eqnarray}
   \cos \alpha &=& \frac{ (m_1 + m_2 ) \cos \frac{\theta_h}{2} }
    { \sqrt{ (m_1 + m_2 )^2 \cos^2 \frac{\theta_h}{2} +
       (m_1 - m_2 )^2 \sin^2 \frac{\theta_h}{2} } },
                                                   \nonumber \\
   \sin \alpha &=& \frac{ - (m_1 - m_2 ) \sin \frac{\theta_h}{2} }
    { \sqrt{ (m_1 + m_2 )^2 \cos^2 \frac{\theta_h}{2} +
      (m_1 - m_2 )^2 \sin^2 \frac{\theta_h}{2} } }.
  \label{bl}
\end{eqnarray}
Using this solution, the ground state energy becomes
\footnote{
The $ m(\theta) $,
$ m(\theta) \equiv
      \sqrt{ (m_1 + m_2 )^2 \cos^2 \frac{\theta}{2} +
        (m_1 - m_2 )^2 \sin^2 \frac{\theta}{2} } $,
is the one which plays an important role in Ref.\cite{rf:MetZhi}.
The $ m(\theta) $ is a monotone decreasing function of $\theta \, ( 0 \le \theta < \pi )$,
   and $ m(\theta=0) = m_1+m_2, \,  m(\theta=\pi) = \vert m_1-m_2 \vert $.
   }
\begin{eqnarray}
  E_{\rm DVV} (\varphi_1, \varphi_2 ) 
    & \approx & -  \vert \langle {\bar q} q \rangle \vert \left( m_1 \cos
         \biggl( \frac{\theta_h}{2} + \alpha \biggr)
            + m_2 \cos \biggl( \frac{\theta_h}{2} - \alpha \biggr) \right)        \nonumber \\
    & = & - \vert \langle {\bar q} q \rangle \vert
            \sqrt{ (m_1 + m_2 )^2 \cos^2 \frac{\theta_h}{2} +
              (m_1 - m_2 )^2 \sin^2 \frac{\theta_h}{2} }.
  \label{bm}
\end{eqnarray}
If one wishes to deal with the parameters $y_1$ and $y_2$ instead of
   $m_1$ and $m_2$, one can represent
\begin{eqnarray}
  E_{\rm DVV} (\varphi_1, \varphi_2 ) 
     &\approx&  
     - \vert \langle {\bar q} q \rangle \vert  \langle S \rangle \sqrt{
      ( y_1 + y_2 )^2 \cos^2 \frac{\theta_h}{2} +  (y_1 - y_2 )^2
       \sin^2 \frac{\theta_h}{2} }              \nonumber  \\
     &\equiv&  - \vert \langle {\bar q} q \rangle \vert \, Y(\theta_h) \, \langle S \rangle,
  \label{bn}
\end{eqnarray}
where
\begin{equation}
    Y(\theta_h) \equiv  \sqrt{ ( y_1 + y_2 )^2 \cos^2 \frac{\theta_h}{2} +  (y_1 - y_2 )^2
       \sin^2 \frac{\theta_h}{2} }.
  \label{bo}
\end{equation}
The $Y(\theta_h)$ is a monotone decreasing function of $\theta_h$ for $0 \le \theta_h <\pi$,
   and \\ $Y(\theta_h=0)=y_1 +y_2, \, Y(\theta_h=\pi)= \vert y_1 - y_2 \vert $.

Let us consider the degenerate case $m_1 =m_2 \equiv m$.
In this case one has $\sin \alpha =0$ and $\cos \alpha =1$ because of the
   condition, $ -\pi < \theta_h < \pi $,
and a solution $\alpha =0$ lesds to
   $ \varphi_1 = \varphi_2 \equiv \varphi $.
 The ground state energy in the degenerate case is then
\begin{eqnarray}
  E_{\rm DVV} (\varphi ) 
     &\approx&  
     - 2 m \vert \langle {\bar q} q \rangle \vert \cos \frac{\theta_h}{2}
     + O  \biggl(  \biggl( \frac{ \vert \langle {\bar q} q \rangle \vert m_j}{\tau}
                               \biggr)^2 \biggr),
   \label{bp}
\end{eqnarray}
which becomes zero if $\theta_h = \pi$.
If we solve the equation of motion for $\varphi$ up to
   $O ( {\vert \langle {\bar q} q \rangle \vert m} / \tau ) $,
   we get
\begin{equation}
       \varphi = \frac{\theta_h}{2}
        - \frac{1}{2}  \biggl( \frac{ \vert \langle {\bar q} q \rangle \vert m}{\tau}\biggr)
           \sin \frac{\theta_h}{2}  
            + O  \biggl(  \biggl( \frac{ \vert \langle {\bar q} q \rangle \vert m_j}{\tau}
                               \biggr)^2 \biggr),
  \label{bq}
\end{equation}
and
\begin{eqnarray}
  E_{\rm DVV} (\varphi ) 
    & \approx & -  2 m \vert \langle {\bar q} q \rangle \vert  \cos \frac{\theta_h}{2}
      - \frac{\tau}{2} \biggl( \frac{ \vert \langle {\bar q} q \rangle \vert m}{\tau}\biggr)^2
         \sin^2 \frac{\theta_h}{2} 
       + O  \biggl(  \biggl( \frac{ \vert \langle {\bar q} q \rangle \vert m_j}{\tau}
                               \biggr)^3 \biggr),
  \label{br}
\end{eqnarray}
which does not become zero when $\theta_h = \pi$.
In the degenerate case if one restricts $\cos { (\theta_h / 2) }$ as
\begin{equation}
    \frac{ m \vert \langle {\bar q} q \rangle \vert }{\tau} \ll  \cos \frac{\theta_h}{2},
  \label{bs}
\end{equation}
the second higher term in the ground state energy $ E_{\rm DVV} $ can be neglected.
\section{Physical effects of the vacuum angle $\theta_h$}
\subsection{On electroweak symmetry breaking \\
                          ---specially on Higgs field---}
In the preceding section we see that the ground state energy for a strongly interacting
   hidden sector, $ E_{\rm DVV} $, Eq.(\ref{bn}), depends on the vacuum angle $\theta_h$.
Since a linear term of $S$ in this $ E_{\rm DVV} $ leads to VEV
   of $S$\cite{rf:HurKo,rf:HolKubLimLin}, the vacuum angle $\theta_h$ will affect EW
   symmetry breaking which is triggered by this $S$.
 When $y_1 \ne y_2$, the tree level scalar potential is
\footnote{
    When $\theta_h=0$, this potential is equal to the one
     in Ref.\cite{rf:HurKo}.
     }
\begin{eqnarray}
   V &=&  V_{ {\rm SM}+ S } + V_{\rm DVV}         \nonumber  \\
      & \approx &  \lambda_H ( H^\dagger H )^2
          + \frac{1}{4} \lambda_S S^4 - \frac{1}{2}  \lambda_{HS} \, S^2 ( H^\dagger H )
           - \vert \langle {\bar q} q \rangle \vert \, Y(\theta_h) \,  S.
  \label{ca}
\end{eqnarray}
With $ H =(0, h/ \sqrt{2})^T $, a linear term in a real singlet scalar $S$ leads to VEV of $S$,
   and then to VEV of the Higgs field $h$\cite{rf:HolKubLimLin},
\begin{eqnarray}
   \langle S \rangle &=&  \biggl[ \biggl(
        \frac{4 \lambda_H}{4  \lambda_H  \lambda_S - \lambda_{HS}^2} \biggr)
        \left\{ \, \vert \langle {\bar q} q \rangle \vert \, Y(\theta_h) \, \right\} \biggr]^\frac{1}{3}
                                                                                             \nonumber  \\  
     &=&  \biggl(
        \frac{4 \lambda_H}{4  \lambda_H  \lambda_S - \lambda_{HS}^2} \biggr)^\frac{1}{3}
           \vert \langle {\bar q} q \rangle \vert^\frac{1}{3}
            \biggl\{ \, ( y_1 + y_2 )^2 \cos^2 \frac{\theta_h}{2} +  (y_1 - y_2 )^2
                         \sin^2 \frac{\theta_h}{2} \, \biggr\}^\frac{1}{6},           \nonumber  \\
      \langle h \rangle &=&
             \langle S \rangle \sqrt{ \frac{\lambda_{HS}}{2 \lambda_H} },
  \label{cb}
\end{eqnarray}
where $ Y(\theta_h) $ has been defined in Eq.(\ref{bo}).
Note that $ \langle h \rangle$ depends on the vacuum angle $\theta_h$ as well
   as $\langle S \rangle$ does.
On the other hand, the $ \langle h \rangle$ should be $246 {\rm GeV}$.
Therefore, the six parameters,
   $( \lambda_H, \lambda_S, \lambda_{HS}, \vert \langle {\bar q} q \rangle \vert, y_1, y_2 )$,
   appeared in Eq.(\ref{cb})
   should satisfy the condition, $ \langle h \rangle =246 {\rm GeV}$, for a given value
   of $\theta_h$.
Since $Y(\theta_h)$ has different values for different values of $\theta_h$,
   the set of the six parameters
    $( \lambda_H, \lambda_S, \lambda_{HS}, \vert \langle {\bar q} q \rangle \vert, y_1, y_2 )$
    for a $\theta_h$ is different from that for another  $\theta_h' (\ne \theta_h)$.
If we first give the values of $  \vert \langle {\bar q} q \rangle \vert, \, y_1,$ and $y_2$,
   the three parameters, $( \lambda_H,  \lambda_S, \lambda_{HS} )$,
   become the remaining
   parameters which are determined for a given value of $\theta_h$ by the condition,
    $ \langle h \rangle =246 {\rm GeV}$.
This means that for different values of $\theta_h$, we take different values
   of $( \lambda_H,  \lambda_S, \lambda_{HS} )$.
In the degenerate case $m_1 =m_2 \equiv m$, or $y_1 =y_2 \equiv y$, though, we should
   pay attention when we use the expression, Eq.(\ref{cb}).
The VEV's, $ \langle S \rangle $ and $ \langle h \rangle $, will be given in the same form,
   Eq.(\ref{cb}), as in the case of $m_1 \ne m_2$ if we can use the form of the tree level
   scalar potential, Eq.(\ref{ca}), for the degenerate case.
In order for the tree level scalar potential to be represented by Eq.(\ref{ca}) also in the
   degenerate case, the second term in the right-hand side of
   Eq.(\ref{br}) should be neglected.
So as to be neglected that second term, we need to restrict the vacuum angle as discussed
   in Eq.(\ref{bs}),
   $  { m \vert \langle {\bar q} q \rangle \vert } / {\tau}  \ll  \cos (\theta_h / 2) $.
The factor $ \cos (\theta_h / 2) $ will not be so small unless the value of
   $ { m \vert \langle {\bar q} q \rangle \vert } / {\tau} $ is nearly zero.

Next, we shall examine whether the condition which determines the Higgs particle mass
   depends on the vacuum angle $\theta_h$ or not.
With $ H =(0, (  \langle h  \rangle +{\hat h} ) / \sqrt{2})^T $ and
   $ S=  \langle S  \rangle +{\hat S} $, the potential $ V = V_{ {\rm SM}+ S } + V_{\rm DVV} $
   leads to the mass matrix,
\begin{eqnarray}
   & &   \frac{1}{2} \left(   \begin{array}{cc}
                     {\hat h}   \, &    \,    {\hat S}          
                                         \end{array}
                                                        \right)
            \left(   \begin{array}{cc}
                     3 \lambda_H  \langle h  \rangle^2 -  \frac{1}{2} \lambda_{HS}
                                                                    \langle S \rangle^2               
                    & \hskip0.3cm
                     - \lambda_{HS}   \langle h  \rangle  \langle S \rangle                   \\
                      - \lambda_{HS}   \langle h  \rangle  \langle S \rangle
                     & \hskip0.9cm
                     3 \lambda_S  \langle S  \rangle^2 -  \frac{1}{2} \lambda_{HS}
                                                                    \langle h \rangle^2
                                         \end{array}
                                                        \right)
            \left(   \begin{array}{c}
                     {\hat h}        \\  
                     {\hat S}                   \end{array}
                                                        \right)                   \nonumber \\
  & = &  \frac{1}{2} \left(   \begin{array}{cc}
                     {\hat h}   \, &    \,    {\hat S}          
                                         \end{array}
                                                        \right)
               \langle h  \rangle^2 
            \left(   \begin{array}{cc}
                     2 \lambda_H     
                    & \hskip0.3cm
                       - \sqrt{2 \, \lambda_H \, \lambda_{HS} }                    \\
                      - \sqrt{2 \, \lambda_H \, \lambda_{HS} }
                     & \hskip0.9cm
                      \frac{6 \, \lambda_S \lambda_H}{\lambda_{HS} } - \frac{\lambda_{HS} }{2}
                                         \end{array}
                                                        \right)
            \left(   \begin{array}{c}
                     {\hat h}        \\  
                     {\hat S}                   \end{array}
                                                        \right),
  \label{cc}
\end{eqnarray}
where we have used
   $ {\langle h \rangle} / { \langle S \rangle} =
             \sqrt{ {\lambda_{HS}} / {2 \lambda_H} }$.
Diagonalizing the mass matrix, one must obtain the mass eigenstate of the Higgs particle
  whose eigenvalue should be $125.1 {\rm GeV}$.
Namely, the matrix,
\begin{equation}
   \left(   \begin{array}{cc}
                     2 \lambda_H     
                    & \hskip0.3cm
                       - \sqrt{2 \, \lambda_H \, \lambda_{HS} }                    \\
                      - \sqrt{2 \, \lambda_H \, \lambda_{HS} }
                     & \hskip0.9cm
                      \frac{6 \, \lambda_S \lambda_H}{\lambda_{HS} } - \frac{\lambda_{HS} }{2}
                                         \end{array}
                                                        \right),
            \label{cd}
\end{equation}
should give the eigenvalue
   ${ (125.1 {\rm GeV})^2 } / { \langle h  \rangle^2} = 0.2586$ of
    the mass eigenstate of the Higgs particle.
This condition does not depend on $\theta_h$, and
   the mixing angle between $\hat h$ and $\hat S$ also does not.
If, however, one involves the next leading order correction,
   $O ( ( { \vert \langle {\bar q} q \rangle \vert m_j} / {\tau} )^2 )$,
   in the ground state energy $E_{\rm DVV}$, the mass matrix will depend on $\theta_h$.
\subsection{On mass difference between dark matter $  \pi^0 $ and $  \pi^{\pm} $}
We will study how the vacuum angle $\theta_h$ contributes to mass difference between
    the hidden-sector pions $ \pi^0 $ and $ \pi^\pm $, which are dark mater candidates.
The masses of these hidden-sector pions $ \pi^0 $ and $ \pi^\pm $ can be obtained using
   the effective theory $ {\cal L}_{\rm DVV} $, Eq.(\ref{bc}), as follows.
Since both fields $\eta^0$ and $ \pi_3 $ have VEV's, Eq.(\ref{bf}), provided $\theta_h \ne 0$,
   it is convenient to
   introduce new fields that have no  VEV's.
We shall define $\hat U$ by
\begin{equation}
   \hat U \equiv U U_g^{-1},
  \label{ce}
\end{equation}
and the fields $ {\hat \pi}_j $ and ${\hat \eta}$ by
\begin{equation}
   \hat U = \exp \biggl[  \frac{i}{f_\pi} \biggl\{ \sum_{j=1}^{3} {\hat \pi}_j \sigma^j
           + {\hat \eta} \cdot \bf 1 \biggr\}  \biggr],
  \label{cf}
\end{equation}
where $U_g$ is defined in Eq.(\ref{bf}).
These new fields $ {\hat \pi}_j $ and ${\hat \eta}$ have no VEV's and
   the Lagrangian $ {\cal L}_{\rm DVV}$ can be rewritten as
   follows with the help of Eq.(\ref{bh})\cite{rf:RosSchTra,rf:VecVen,rf:KawOht,rf:NatArn},
\begin{eqnarray}
  & &   \hskip-0.7cm    {\cal L}_{\rm DVV}                                       \nonumber \\
  &=& \frac{f_\pi^2}{4} \, {\rm Tr} \{ \partial_\mu {\hat U} \partial^\mu {\hat U}^\dagger \}
          + \frac{ \vert \langle {\bar q} q \rangle \vert }{2}  \, 
                {\rm Tr} \{ {\cal M}(\theta_h) ( {\hat U} + {\hat U}^\dagger - 2 ) \}   
               + \frac{\tau}{8} \biggl\{ \log \det {\hat U} - \log \det {\hat U}^\dagger      
            \biggr\}^2                                                   \nonumber \\
   & & \hskip0.1cm  - i \, \frac{\tau}{2} \{ \theta_h - (\varphi_1 + \varphi_2) \} \, {\rm Tr}
           \biggl\{ \log \frac{\hat U}{ {\hat U}^\dagger } - ( {\hat U} - {\hat U}^\dagger )
                                              \biggr\} + {\rm const.},
  \label{cg}
\end{eqnarray}
where
   $ {\cal M}(\theta_h) \equiv {\rm diag.} ( m_1 \cos \varphi_1, \, m_2 \cos \varphi_2) $.
Expanding the exponential in $\hat U$, we get
\begin{eqnarray}
     & &  \hskip-1.5cm
     \frac{f_\pi^2}{4} \, {\rm Tr} \{ \partial_\mu {\hat U} \partial^\mu {\hat U}^\dagger \}
          + \frac{ \vert \langle {\bar q} q \rangle \vert }{2}  \, 
                {\rm Tr} \{ {\cal M}(\theta_h) ( {\hat U} + {\hat U}^\dagger - 2 ) \}   
               + \frac{\tau}{8} \biggl\{ \log \det {\hat U} - \log \det {\hat U}^\dagger \biggr\}^2
                                                                                                    \nonumber \\
      & \approx & \frac{1}{2} ( \partial_\mu {\hat {\vec \pi}} ) \cdot ( \partial^\mu {\hat {\vec \pi}} )
              +\frac{1}{2}  ( \partial_\mu {\hat \eta} )( \partial^\mu {\hat \eta} )
              - \frac{1}{2} \biggl\{ \frac{  m_+(\theta_h)
                   \vert \langle {\bar q} q \rangle \vert }{f_\pi^2}
                    \biggr\} \, ( {\hat {\vec \pi}} \cdot {\hat {\vec \pi}} )                       \nonumber \\
       & & \hskip0.3cm 
       -  \frac{1}{2} \biggl\{ \frac{4 \, \tau}{f_\pi^2} 
          + \frac{  m_+(\theta_h)
          \vert \langle {\bar q} q \rangle \vert }{f_\pi^2} \biggr\} {\hat \eta}^2
          - \frac{1}{2}  \biggl\{  \frac{ 2 \, m_-(\theta_h) 
                \vert \langle {\bar q} q \rangle \vert }{f_\pi^2} \biggr\}
                 ( {\hat \pi}_3 \cdot {\hat \eta})+\cdots                                          \nonumber \\
        &=&  \frac{1}{2} ( \partial_\mu {\hat {\vec \pi}} ) \cdot ( \partial^\mu {\hat {\vec \pi}} )
                 +\frac{1}{2} ( \partial_\mu {\hat \eta} )( \partial^\mu {\hat \eta} )
                 -  \frac{1}{2} \biggl\{
               \frac{ \vert \langle {\bar q} q \rangle \vert m_+(\theta_h)}{f_\pi^2} \biggr\} \,
                ( {\hat \pi}_1^2 + {\hat \pi}_2^2 )
                                                                                 \nonumber \\
        & & \hskip0.3cm - \frac{1}{2}
           \left(   \begin{array}{cc}
                     {\hat \eta}   \, &    \,    {\hat \pi}_3          
                                         \end{array}
                                                        \right)
         \left(   \begin{array}{cc}
        \frac{4 \, \tau}{f_\pi^2}+ \frac{ \vert \langle {\bar q} q \rangle \vert m_+(\theta_h)} {f_\pi^2}
                    & \hskip0.5cm
                    \frac{ \vert \langle {\bar q} q \rangle \vert  m_-(\theta_h) }{f_\pi^2}
                                                                                                                                 \\
                      \frac{ \vert \langle {\bar q} q \rangle \vert  m_-(\theta_h) }{f_\pi^2}         
                     & \hskip0.5cm
                     \frac{ \vert \langle {\bar q} q \rangle \vert m_+(\theta_h) }{f_\pi^2}   
                                         \end{array}
                                                        \right)
            \left(   \begin{array}{c}
                     {\hat \eta}        \\  
                     {\hat \pi}_3                   \end{array}
                                                        \right) +\cdots,
  \label{ch}
\end{eqnarray}
where we have defined
\begin{eqnarray}
    m_+(\theta_h) & \equiv & m_1 \cos \varphi_1 +m_2 \cos \varphi_2,         \nonumber \\
    m_-(\theta_h) & \equiv & m_1 \cos \varphi_1 -m_2 \cos \varphi_2.
  \label{ci}
\end{eqnarray}
In the case of $m_1 \ne m_2$, we have
   $ m_-(\theta_h) = { (m_1^2 - m_2^2) } / { m_+(\theta_h) } $
   because of $ m_2  \sin \varphi_2 = m_1  \sin \varphi_1 $.
We see that masses of the hidden-sector pions and flavor-singlet pseudoscalar depend
   on the hidden-sector vacuum angle $\theta_h$
   \cite{rf:VecVen,rf:MetZhi,rf:AntRedStrVig}.
Note that the factor $ m_+(\theta_h)$ in the diagonal term of ${\hat \pi}_3$ appears in
   the right-hand side of Eq.(\ref{bm}) and is a monotone decreasing function of
   $\theta_h$ ($0 \le \theta_h \le \pi$),
   while the absolute value of the factor $ m_-(\theta_h)$ in the mixing term
   between ${\hat \pi}_3$ and $\hat \eta$ is a monotone increasing function of $\theta_h$.
The mass eigenstates $\hat \eta'$ and ${\hat \pi}^0$,
\begin{equation}
   \left(   \begin{array}{c}
                     {\hat \eta'}        \\  
                     {\hat \pi}^0           \end{array}
                                                        \right) 
   = \left(   \begin{array}{cc}
                   \cos \alpha           
                    & \hskip0.3cm
                   - \sin \alpha                    \\
                     \sin \alpha
                     & \hskip0.6cm
                    \cos \alpha
                                         \end{array}
                                                        \right)
            \left(   \begin{array}{c}
                     {\hat \eta}        \\  
                     {{\hat \pi}_3}                   \end{array}
                                                        \right),
             \label{cj}
\end{equation}
can be obtained by diagonalizing the mass matrix and resultant mixing angle $\alpha$
   is small,
   $ \vert \sin \alpha \vert \approx \vert \langle {\bar q} q \rangle \vert \,
        \vert m_-(\theta_h) \vert / (4 \tau)  \ll 1 $.
One obtains the mass eigenstate $\hat \eta'$ with the eigenvalue(squared),
\begin{equation}
    m( \eta' )^2 =  \biggl\{ \frac{4 \, \tau}{f_\pi^2} 
          + \frac{ \vert \langle {\bar q} q \rangle \vert m_+(\theta_h) }{f_\pi^2} \biggr\} 
          +  \frac{ \vert \langle {\bar q} q \rangle \vert^2 m_-(\theta_h)^2 }{4 \, \tau f_\pi^2},
     \label{ck}
\end{equation}
and the mass eigenstate ${\hat \pi}^0$ with the eigenvalue(squared),
\begin{equation}
      m( \pi^0 )^2  = 
         \biggl\{ \frac{ \vert \langle {\bar q} q \rangle \vert m_+(\theta_h)  }{f_\pi^2} \biggr\} 
          -  \frac{ \vert \langle {\bar q} q \rangle \vert^2  m_-(\theta_h)^2}{4 \, \tau f_\pi^2}.
    \label{cl}
\end{equation}

The mass difference between the hidden-sector pions
   ${\hat \pi}^0$ and ${\hat \pi}^{\pm}$ can be represented,
\begin{eqnarray}
   \left. \frac{ m( \pi^{\pm} )^2 - m( \pi^0 )^2 }{ m( \pi^{\pm} )^2 } \right|_{\theta_h}
   & = &  \frac{ \vert \langle {\bar q} q \rangle \vert}{4 \, \tau}
               \frac{m_-(\theta_h)^2}{m_+(\theta_h)}                   \nonumber \\
   & \approx &   \frac{ \vert \langle {\bar q} q \rangle \vert}{4 \, \tau}
               \frac{ (m_1 + m_2)^2 (m_1 - m_2)^2 }
               { \biggl\{  (m_1 + m_2 )^2 \cos^2 \frac{\theta_h}{2}
                         +  (m_1 - m_2 )^2 \sin^2 \frac{\theta_h}{2} \biggr\}^\frac{3}{2}  }.
  \label{cm}
\end{eqnarray}
Note that the right-hand side
   of the above equation is a monotone increasing function of $\theta_h$
   ($0 \le \theta_h \le \pi$),
   then the mass difference between ${\hat \pi}^0$ and ${\hat \pi}^{\pm}$
   becomes large as the $\theta_h$ changes from $0$ to $\pi$.
When $\theta_h=0$, the right-hand side of the above equation
   is very small,
\begin{equation}
   \left.
   \frac{ m( \pi^{\pm} )^2 - m( \pi^0 )^2 }{ m( \pi^{\pm} )^2 } \right|_{\theta_h=0}
   \approx \frac{ \vert \langle {\bar q} q \rangle \vert (m_1 +m_2)}{ 4 \tau} 
      \times \biggl( \frac{m_1 -m_2}{m_1 +m_2} \biggr)^2  \ll  1,
  \label{cn}
\end{equation}
thereby ${\hat \pi}^0$ and ${\hat \pi}^{\pm}$ are almost degenerate.
However, when $\theta_h=\pi$, the mass difference will not be so small,
\begin{equation}
   \left. \frac{ m( \pi^{\pm} )^2 - m( \pi^0 )^2 }{ m( \pi^{\pm} )^2 } \right|_{\theta_h=\pi}
   \approx  \frac{ \vert \langle {\bar q} q \rangle \vert (m_1 +m_2)}{ 4 \tau} 
      \times \biggl( \frac{m_1 +m_2}{ \vert m_1 -m_2 \vert } \biggr).
  \label{co}
\end{equation}
If the value of $ (m_1 +m_2) / ( \vert m_1 -m_2 \vert ) $ (the inverse of the hidden-sector
   isospin violation
   in $\theta_h=0$ ) is rather large, the mass difference between ${\hat \pi}^0$ and 
   ${\hat \pi}^{\pm}$ can not be neglected.
Among the four particles,
   $({\hat \pi}^0, {\hat \pi}^+, {\hat \pi}^-, {\hat \eta'} )$, ${\hat \pi}^0$ is most
   light, while ${\hat \eta'}$ is most heavy\cite{rf:Wit,rf:VicPan}.
Since ${\hat \pi}^0, {\hat \pi}^+$, and ${\hat \pi}^-$ are stable ($ {\hat \pi}^+$ and $ {\hat \pi}^-$
   have the Cartan subalgebra $U(1)$ charge +1 and -1,
   respectively when $y_1 \ne y_2$ \cite{rf:AmeAokGotKub}), they are dark matters.

In the degenerate case, $m_1 = m_2 \equiv m$, we have restricted $\theta_h$ in Sec.3.1
   as \\
   $  { m \vert \langle {\bar q} q \rangle \vert } / {\tau}  \ll  \cos (\theta_h / 2) $.
The coefficient of the mixing term $ ( {\hat \pi}_3 \cdot {\hat \eta})$ in Eq.(\ref{ch})
   vanishes when $ m_1 = m_2 $ as
   $ m_-(\theta_h) = m \cos \varphi - m \cos \varphi = 0 $
   and then all the three pions have the same mass,
\begin{eqnarray}
   \frac{ \vert \langle {\bar q} q \rangle \vert m_+(\theta_h)  }{f_\pi^2}
    & \approx &  \frac{ 2 m \vert \langle {\bar q} q \rangle \vert }{f_\pi^2} \cos \frac{\theta_h}{2},
  \label{cp}
\end{eqnarray}
depending on the hidden-sector vacuum angle $\theta_h$ \cite{rf:MetZhi}.\\
\\
\section{Conclusion}

Considering vacuum angle $\theta_h$ in a strongly interacting hidden sector like QCD,
   we studied how the vacuum angle $\theta_h$ affects physical quantities in a scale
   invariant extension of the standard model with that hidden sector proposed by
   Hur and Ko\cite{rf:HurKo}.
In their model, the dynamical chiral symmetry breaking occurs in the hidden
   sector and generates a scale $ \vert \langle {\bar q} q \rangle \vert $ which can trigger
   electroweak symmetry breaking.
As the low-energy effective theory of the hidden sector with $\theta_h$ like QCD, we have
   used the Di-Vecchia-Veneziano model with two flavors.

We find explicitly how the expression for VEV of the Higgs field which should be
   $246 {\rm GeV}$
   depends on the vacuum angle $\theta_h$ besides the parameters in our model.
On the other hand, it is shown that
   the condition which determines the Higgs particle mass does not depend
   on $\theta_h$ in a first approximation of small hidden-sector quark masses,
   $m_1$ and $m_2$.
Next, we studied the mass difference between the hidden-sector pions,
   the hidden-sector isotriplet pions $ {\vec \pi} =(  \pi_1, \pi_2, \pi_3 ) $ being candidates
   for cold dark matter.
The arising mixing term between $\pi_3$ and a $SU(2)$ flavor-singlet $\eta^0$ depends on
   $\theta_h$, and this term leads to the mass difference between dark matter
   $\pi^0$ and $\pi^{\pm}$.
We find that while the mass difference between $\pi^0$ and $\pi^{\pm}$ is negligible
   for $\theta_h \approx 0$, the mass difference will not be so small for $\theta_h \approx \pi$
   when $ (m_1 +m_2) / \vert m_1 -m_2 \vert $ is rather large.

Since no hidden-sector particle is detected in experiment, the vacuum angle $\theta_h$ in
   the strongly interacting hidden sector is not restricted.
It is therefore worthwhile to study the scale invariant extension of
    the standard model\cite{rf:HurKo} with any value of  $\theta_h$.
While we have restricted ourselves to the two flavor case, $N_{h,f}=2$, of the hidden sector,
   it will be interesting to study the case of $N_{h,f} \ge 3 $.
%
%
%
%
%
%
%
%
%
%
%
\newpage
\end{document}